\documentstyle[preprint,amsfonts,aps]{revtex}

\newcommand{\sous}[2]{\stackrel{\phantom{(n,p)}}{#2} \! \! \! \! \! \!
			\! \! \! 
			{{} \atop \scriptstyle #1} {}}

\begin{document}
\draft
\title{On the equations of motion of point-particle binaries \\ 
at the third post-Newtonian order}
\author{Luc Blanchet$^{1,2}$ and Guillaume Faye$^1$}
\address{$^1$ D\'epartement d'Astrophysique Relativiste et de
Cosmologie (CNRS), \\ 
Observatoire de Paris, 92195 Meudon Cedex, France}
\address{$^2$ Max-Planck-Institut f\"ur Gravitationsphysik, 
Albert-Einstein-Institute, \\ Am M\"uhlenberg 1, D-14476 Golm, Germany}
\date{\today}

\maketitle
\widetext
\begin{abstract}
We investigate the dynamics of two point-like particles through the
third post-Newtonian (3PN) approximation of general relativity. The
infinite self-field of each point-mass is regularized by means of
Hadamard's concept of ``partie finie''. Distributional forms
associated with the regularization are used systematically in the
computation. We determine the stress-energy tensor of point-like
particles compatible with the previous regularization. The Einstein
field equations in harmonic coordinates are iterated to the 3PN
order. The 3PN equations of motion are Lorentz-invariant and admit a
conserved energy (neglecting the 2.5PN radiation reaction). They
depend on an undetermined coefficient, in agreement with an earlier
result of Jaranowski and Sch\"afer. This suggests an incompleteness of
the formalism (in this stage of development) at the 3PN order. In this
paper we present the equations of motion in the center-of-mass frame
and in the case of circular orbits.
\end{abstract}


\narrowtext

A cardinal problem in Gravitational Physics is that of the dynamics of
binary systems of point particles. In general relativity, this problem
is tackled by means of the post-Newtonian approximation, or formal
expansion when the speed of light $c$ goes to infinity. By definition,
the $n$PN approximation refers to the terms in the equations of motion
that are smaller than the Newtonian force by a factor of order
$1/c^{2n}$. For the motion of two non-spinning point particles, the
1PN approximation was obtained first by Lorentz and Droste
\cite{LD17}. Subsequently, Einstein, Infeld and Hoffmann \cite{EIH}
re-derived the 1PN order using their famous ``surface-integral''
method. In the eighties, Damour and Deruelle \cite{DD}, starting from
a ``post-Minkowskian'' iteration scheme developed by Bel {\it et al}
\cite{BeDD81}, were able to compute the equations of motion up to the
2.5PN order, at which the gravitational-radiation reaction effects
first take place. The motivation was to firmly establishing the rate
at which the orbit of the binary pulsar PSR 1913+16 decays because of
gravitational-radiation emission. The 2.5PN approximation was then
obtained by Sch\"afer \cite{S} using an ``ADM Hamiltonian'' approach
initiated by Ohta {\it et al} \cite{OO}. Furthermore, Kopeikin {\it et
al} \cite{Kop} derived the same result within their ``extended-body''
method (without any need of a regularization). More recently, the
2.5PN equations of motion as well as 2.5PN gravitational field were
derived by Blanchet, Faye and Ponsot \cite{BFP98} applying a direct
``post-Newtonian'' iteration of the field equations. Finally,
Jaranowski and Sch\"afer \cite{JaraS98} investigated within the
Hamiltonian approach the 3PN order and found some ambiguities linked
to the regularization of the self-field of point masses. As for them,
the 3.5PN terms in the equations of motion are well-known \cite{reac};
they are associated with higher-order radiation reaction effects.

The motivation for working out the 3PN equations of motion is not the
timing of the binary pulsar anymore, but the detection of
gravitational radiation by future experiments such as LIGO and
VIRGO. Indeed, the 3PN equations are needed in particular to construct
accurate 3.5PN templates for detecting and analyzing the waves
generated by inspiralling compact binaries \cite{data}. Currently, we
know the complete templates up to the 2.5PN order \cite{BDIWW}, plus
the contribution therein of non-linear effects to the 3.5PN order
\cite{B98tail}, plus all the terms in the vanishing mass-ratio limit
to the 5.5PN order \cite{TTS}. In this Letter, we outline our
derivation of the 3PN equations of motion, based on the direct
post-Newtonian approach of \cite{BFP98}, and we present the result in
the case, appropriate to inspiralling compact binaries, of circular
orbits. We confirm by means of a well-defined regularization {\it \`a la}
Hadamard the finding of Jaranowski and Sch\"afer \cite{JaraS98} that
there remains at the 3PN order an undetermined coefficient appearing
in front of a quartically non-linear term (proportional to $G^4$).

Consider the class ${\cal F}$ of functions $F({\bf x})$ that are
smooth on ${\mathbb R}^3$ except at two isolated points ${\bf y}_1$
and ${\bf y}_2$, around which they admit a power-like singular
expansion of the form

\begin{equation}\label{1}
\forall n\in {\mathbb N}\;,\quad
F({\bf x})=\sum_{a_0\leq a\leq n} r_1^a \!\!\sous{1}{f}_{a}({\bf
n}_1)+o(r_1^n)\quad\hbox{when $r_1\to 0$} \;,
\end{equation}
where $r_1=|{\bf x}-{\bf y}_1|$, ${\bf n}_1=({\bf x}-{\bf y}_1)/r_1$,
and where the powers $a$ are supposed to be real, to range in discrete
steps: $a\in (a_i)_{i\in {\mathbb N}}$, and to be bounded from below:
$a_0\leq a$. The coefficients ${}_1f_a$ of the various powers of $r_1$
in this expansion are smooth functions of the unit vector ${\bf
n}_1$. We refer to the coefficients ${}_1f_a$ with $a<0$ as the {\it
singular} coefficients of $F$ around 1; their number is always
finite. Moreover, we have the same type of expansion around the other
point (when $r_2\to 0$). The Hadamard ``partie finie'' \cite{Hadamard}
of $F$ at the location of the singular point 1 is equal to the angular
average of the zeroth-order coefficient in (\ref{1}), i.e.

\begin{equation}\label{2}
(F)_1= \int {d\Omega_1\over 4\pi} \!\!\sous{1}{f}_0({\bf n}_1) \;,
\end{equation}
with $d\Omega_1= d\Omega ({\bf n}_1)$ the usual solid angle
element. The partie finie is ``non-distributive'' in the sense that
$(FG)_1\not= (F)_1(G)_1$ in general. Besides (\ref{2}), we define also
the partie finie (${\rm Pf}$) of the divergent integral $\int d^3{\bf
x}~F$, assuming that $F$ decreases sufficiently rapidly when $|{\bf
x}|\to +\infty$ so that the divergencies come only from the singular
points 1 and 2. With full generality \cite{Hadamard,Schwartz},

\begin{eqnarray}\label{3}
{\rm Pf}_{s_1,s_2}\int d^3{\bf x}~ F &=&~\lim_{s\to
0}~\biggl\{\int_{{\mathbb R}^3\setminus {\cal B}_1(s)\cup {\cal
B}_2(s)}d^3{\bf x}~ F\nonumber\\ &&\qquad +~4\pi\sum_{a+3<
0}{s^{a+3}\over a+3} \left({F\over r_1^a}\right)_1+ 4 \pi
\ln\left({s\over s_1}\right) \left(r_1^3 F\right)_1 +1\leftrightarrow
2\biggr\}\;.
\end{eqnarray}
The first term is the finite integral over ${\mathbb R}^3$ deprived
from the two spherical balls ${\cal B}_1(s)$ and ${\cal B}_2(s)$ with
radius $s$ and centred on the two singularities. The extra terms are
such that they exactly cancel out the divergent part of the integral
when $s\to 0$ (the notation $1\leftrightarrow 2$ indicates the same
extra terms but referring to the other singularity point). The
logarithmic terms depend on two arbitrary positive constants $s_1$ and
$s_2$ that come from the arbitrariness in the choice of unit length
for measuring $s$; hence, the partie finie depends on both $s_1$ and
$s_2$ (as indicated by the notation ${\rm Pf}_{s_1,s_2}$). Applying
(\ref{3}) to the case of a gradient $\partial_i F$, we find
\cite{BFreg}

\begin{equation}\label{4}
{\rm Pf}\int d^3{\bf x}~ \partial_i F = -4 \pi (n_1^i r_1^2 F)_1 
+1\leftrightarrow 2\;.
\end{equation}
In words, the integral of a gradient is equal to the sum of the
surface integrals surrounding the two singularities, in the limit
where the surface areas shrink to zero and following the
regularization (\ref{2}). Thus, the integral of a gradient is not zero
in general, which shows that the ``ordinary'' derivative $\partial_i
F$ is not adequate for applying to point-particles a formalism
initially valid for continuous sources, since in the latter case the
integral of a gradient does never contribute. To define a ``better''
notion of derivative, we must construct the distributional forms
associated with the functions in the class ${\cal F}$.

For any $F\in {\cal F}$, we consider the ``pseudo-function'' ${\rm Pf}
F$ defined as the linear form on ${\cal F}$ such that $\forall G \in
{\cal F}$, $< {\rm Pf} F, G > ={\rm Pf} \int d^3{\bf x}~ F G$, the
duality bracket denoting here the result of the action of ${\rm Pf} F$
on the function $G$. The product of pseudo-functions is defined to be
the ordinary pointwise product that we use in Physics, i.e. ${\rm Pf}
F~\!\!.~\!{\rm Pf} G={\rm Pf} (FG)$.  With the help of the Riesz
\cite{Riesz} delta-function $\delta_\varepsilon({\bf x}-{\bf
y}_1)=\case{\varepsilon (1-\varepsilon)}{4\pi}~\!r_1^{\varepsilon-3}$,
which belongs to the class ${\cal F}$, we construct \cite{BFreg} the
pseudo-function ${\rm Pf}\delta_1$ (in the limit $\varepsilon\to 0$);
by definition: $\forall F\in {\cal F}$, $< {\rm Pf}\delta_1 , F> =
(F)_1$. Clearly ${\rm Pf}\delta_1$ generalizes the standard Dirac
distribution $\delta_1\equiv \delta ({\bf x}-{\bf y}_1)$ to the case
of the Hadamard regularization of the functions in ${\cal
F}$. Furthermore, consistently with the product of pseudo-functions,
we construct the object ${\rm Pf}(F\delta_1)$ which is such that
$\forall G\in {\cal F}$, $< {\rm Pf}(F\delta_1), G> = (FG)_1$. A
trivial consequence of the non-distributivity of the Hadamard partie
finie is that ${\rm Pf}(F \delta_1)\not= (F)_1 {\rm Pf}\delta_1$ in
general cases. The derivative of the pseudo-function ${\rm Pf}F$ is
then obtained from the requirements that (i) the ``rule of integration
by parts'' is satisfied, i.e. $\forall F,G\in {\cal F}$, $<
\partial_i({\rm Pf}F) , G > = -< \partial_i({\rm Pf}G) , F >$, (ii)
the derivative reduces to the ``ordinary'' one in the case where all
the singular coefficients of $F$ vanish. These requirements imply in
particular that $< \partial_i({\rm Pf}F) , 1 >=0$, i.e. the integral
of a gradient is zero. A derivative operator satisfying (i) and (ii)
is given by \cite{BFreg}

\begin{equation}\label{5}
\partial_i ({\rm Pf} F) = {\rm Pf}\biggl( \partial_i F 
+ 4\pi~\! n_1^i \biggl[\case{1}{2}~\!r_1 \!\!\!\sous{1}{f}_{-1}
+\sum_{k\geq 0} {1\over r_1^k} \!\!\!\sous{1}{f}_{-2-k}\biggl] \delta_1 
+1\leftrightarrow 2\biggr)
\end{equation}
(assuming for simplicity that the ${}_1f_a$'s have $a\in {\mathbb
Z}$).  This derivative reduces to the standard distributional
derivative of Schwartz \cite{Schwartz} when applied on smooth
functions with compact support.  We refer to \cite{BFreg} for the
construction of the most general derivative operator satisfying (i),
(ii) and, in addition, (iii) the rule of commutation of derivatives
[not obeyed by (\ref{5})]. One can show however that it does not
satisfy in general the Leibniz rule for the derivative of a
product. The derivative (\ref{5}) is sufficient in the derivation of
the results below. See \cite{BFreg} for details about the Hadamard
regularization and the associated pseudo-functions.

In the post-Newtonian application we are led to consider the partie
finie, in the sense of (\ref{3}), of the Poisson integral of $F$,
i.e. ${\rm Pf}\int d^3{\bf x}~\!F({\bf x})/|{\bf x}-{\bf x}'|$; more
specifically, we are interested in the regularized value, in the sense
of (\ref{2}), of the latter Poisson integral at the location of the
singular point 1, i.e. when $r'_1=|{\bf x}'-{\bf y}_1|\to 0$. We
obtain \cite{BFreg}

\begin{mathletters}\label{6}\begin{eqnarray}
\left({\rm Pf}_{s_1, s_2}  \int \frac{d^3{\bf x}}{|{\bf x}-{\bf x}'|}
~F\right)_1&=&{\rm Pf}_{s_1, s_2} \int \frac{d^3{\bf x}}{r_1}
~F-4\pi\biggl[\ln\biggl({r_1'\over s_1}\biggr)-1\biggr] \bigl(r_1^2
F\bigr)_1 \label{6a}\\ &=&4\pi\ln\biggl({r_{12}\over
r_1'}\biggr)\bigl(r_1^2 F\bigr)_1+4\pi\ln\left({r_{12}\over
s_2}\right)\biggl({r_2^3\over r_1} F\biggr)_2+\cdots\label{6b}
\end{eqnarray}\end{mathletters}$\!\!$
(with $r_{12}=|{\bf y}_1-{\bf y}_2|$).  The first term in (\ref{6a})
represents simply what we get by replacing formally ${\bf x}'$ by
${\bf y}_1$ inside the integrand of the Poisson integral. The second
term is due to the presence of some logarithms $\ln r'_1$ in the
expansion of the integral. (An adaptation of the previous formalism,
detailed in \cite{BFreg}, is needed to take these logarithms into
account, as well as the presence of the integrable singularity ${\bf
x}'$.) As, at last, the $\ln r'_1$ can be gauged away, we regard it as
a constant, taking some finite value (even though $r'_1\to 0$). We check, on
the other hand, that the true constant $s_1$ cancels out between the
two terms of (\ref{6a}), so that the result depends only on $\ln r'_1$
and $\ln s_2$. The complete dependence of the partie finie on these
constants is shown in (\ref{6b}), with the convention that the dots
indicate the terms independent of the constants.

The Einstein field equations relaxed by the condition of harmonic
coordinates [i.e. $\partial_\nu h^{\mu\nu}=0$ with
$h^{\mu\nu}=\sqrt{-g} g^{\mu\nu}-
\eta^{\mu\nu}$; $g={\rm det}g_{\mu\nu}$; $\eta^{\mu\nu}=$diag$(-1,1,1,1)$] read as

\begin{equation}\label{7}
\Box h^{\mu\nu} = \case{16\pi G}{c^4} (-g) T^{\mu\nu}
+\Lambda^{\mu\nu} [h,\partial h,\partial^2h]\;,
\end{equation} 
where $T^{\mu\nu}$ is the matter stress-energy tensor and
$\Lambda^{\mu\nu}$ a complicated functional of $h$ which is at least
of order $O(h^2)$ (and where
$\Box=\eta^{\mu\nu}\partial_\mu\partial_\nu$). We start by
constructing a post-Newtonian solution of (\ref{7}), initially valid
in the case of a continuous (``fluid'') matter tensor $T^{\mu\nu}$,
and parametrized by some appropriate potentials.  We define a
``Newtonian'' potential $V=\Box^{-1}_R [-4\pi G \sigma]$ where
$\Box^{-1}_R$ denotes the standard retarded integral and
$\sigma=(T^{00}+T^{ii})/c^2$; we also introduce a 1PN
``gravitomagnetic'' potential $V_i=\Box^{-1}_R [-4\pi G \sigma_i]$
where $\sigma_i=T^{0i}/c$; some 2PN potentials ${\hat X}$, ${\hat
R}_i$ and ${\hat W}_{ij}$, e.g. ${\hat W}_{ij}=\Box^{-1}_R\left[-4 \pi
G (\sigma_{ij} - \delta_{ij} \sigma_{kk}) - \partial_i V \partial_j
V\right]$ where $\sigma_{ij}=T^{ij}$; and finally some 3PN potentials
${\hat T}$, ${\hat Y}_i$ and ${\hat Z}_{ij}$. In particular, the
potential ${\hat W}_{ij}$ generates the non-linear term $\Box^{-1}_R [
{\hat W}_{ij} \partial_{ij} V ]$, involving a cubic ($G^3$)
contribution, which is part of the potential ${\hat X}$ (many other
cubic terms are contained in ${\hat T}$ and ${\hat Y}_i$). With a
specific choice of potentials we can arrange that all the quartic
($G^4$) terms in the metric appear in ``all-integrated'' form. Since
$V$ is dominantly Newtonian, it needs to be evaluated at the 3PN order
but, for instance, the term $\Box^{-1}_R [ \hat{W}_{ij} \partial_{ij}
V ]$, inside the 2PN potential ${\hat X}$, needs only a relative 1PN
precision. The metric is expressed as a functional of all these
potentials; and with our particular choice of potentials, it turns out
not to be too complicated.

An important point is now to determine the expression of the matter
stress-energy tensor $T^{\mu\nu}$ appropriate to the description of
point-particles. We demand that the dynamics of point-masses follows
from the variation, with respect to the metric, of the action

\begin{equation}\label{7'}
I_{\rm point-particle} = -m_1 c \int_{-\infty}^{+\infty} dt 
\sqrt{-(g_{\mu\nu})_1 v_1^\mu v_1^\nu} + 1\leftrightarrow 2 \;,
\end{equation}
where $v_1^\mu = (c,d {\bf y}_1 /dt)$ is the coordinate velocity of
particle 1. We can check that to the 3PN order all the metric
coefficients $g_{\mu\nu}$ belong to ${\cal F}$ (treating $\ln r'_1$ as
a constant); so $(g_{\mu\nu})_1$ in (\ref{7'}) denotes the value of
the metric at 1 in the sense of (\ref{2}) [or, rather, in the sense of
a Lorentz-covariant Hadamard regularization defined below]. The
stationarity of the action with respect to a metric variation within
the class ${\cal F}$ (i.e. $\delta g_{\mu\nu}\in {\cal F}$) yields the
stress-energy tensor

\begin{equation}\label{8}
T^{\mu\nu}_{\rm point-particle}={m_1 v_1^\mu v_1^\nu \over
\sqrt{-(g_{\rho\sigma})_1 v_1^\rho v_1^\sigma/c^2}} ~\!{\rm
Pf}\left({\delta_1\over \sqrt{-g}}\right) + 1\leftrightarrow 2 \;,
\end{equation}
where the pseudo-function ${\rm Pf}(\case{1}{\sqrt{-g}}\delta_1)$ is
of the type ${\rm Pf}(F\delta_1)$ defined before. [From the rule of
multiplication of pseudo-functions we find that the matter source term
in (\ref{7}) involves the pseudo-function ${\rm
Pf}(\sqrt{-g}\delta_1)$.] To obtain the equations of motion
of the particle 1 we integrate the matter equations of motion
$\nabla_\nu T^{\mu\nu}_{\rm point-particle} = 0$ over a volume
surrounding 1 (exclusively), and use the properties of
pseudo-functions. The equations turn out to have the same form as the
geodesic equations, not with respect to some smooth background but
with respect to the regularized metric generated by the two
bodies. Namely,

\begin{equation}\label{9}
{d\over dt}\left({(g_{\mu\nu})_1 v_1^\nu\over
\sqrt{-(g_{\rho\sigma})_1 v_1^\rho v_1^\sigma}}\right)={1\over 2}
{(\partial_\mu g_{\nu\lambda})_1v_1^\nu v_1^\lambda\over
\sqrt{-(g_{\rho\sigma})_1 v_1^\rho v_1^\sigma}}\;,
\end{equation}
where all the quantities at 1 are evaluated using the
regularization. Let us emphasize that the equations of motion
(\ref{9}) are derived from the specific expression (\ref{8}) of the
stress-energy tensor; had we used another expression, e.g. by
replacing ${\rm Pf}(\case{1}{\sqrt{-g}}\delta_1)\rightarrow
(\case{1}{\sqrt{-g}})_1{\rm Pf}\delta_1$ inside (\ref{8}) (which is
forbidden by the non-distributivity of Hadamard's partie finie), we
would have obtained some different-looking, and {\it a priori}
uncorrect, equations.

The regularization (\ref{2}) is defined {\it stricto sensu} within the
spatial slice $t=$const, and therefore should prevent, at some stage,
the equations of motion from being Lorentz invariant (recall that the
harmonic-gauge condition preserves the Lorentz invariance). It is
known that regularizing within the slice $t=$const yields the correct,
Lorentz-invariant, equations of motion up to the 2PN level
\cite{BFP98}. We find that the breakdown of the Lorentz invariance due
to the regularization (\ref{2}) occurs precisely at the 3PN
order. Therefore, starting at this order, we {\it must} in fact apply
a Lorentz-covariant regularization in (\ref{7'})-(\ref{9}). Evidently,
the good thing to do is to apply the Hadamard regularization in the
frame at which the particle is instantaneously at rest. Let us
consider the Lorentz boost $x'^\mu=\Lambda^\mu_{~\nu}({\bf V})x^\nu$,
where ${\bf V}$ denotes the constant boost velocity. We replace all
the quantities in the original frame by their equivalent expressions,
developed to the 3PN order, in the new frame. Notably, the
trajectories ${\bf y}_1(t)$, ${\bf y}_2(t)$ and velocities ${\bf
v}_1(t)$, ${\bf v}_2(t)$ are replaced by certain functionals of ${\bf
x}'$ and the new trajectories ${\bf y}'_1(t')$, ${\bf y}'_2(t')$ and
velocities ${\bf v}'_1(t')$, ${\bf v}'_2(t')$ (where $t'=x'^0/c$). We
apply the Hadamard regularization within the slice $t'=$const, keeping
${\bf V}$ as a constant ``spectator'' vector. Finally, we re-express
all the quantities back into the original frame at the point 1
($r_1\to 0$), under the condition that ${\bf v}'_1(t')=0$ and
(equivalently) ${\bf V}={\bf v}_1(t)$. This ensures that the new frame
is indeed the rest frame of the particule 1 at the instant $t$. Now
the 3PN equations of motion are Lorentz invariant.
 
All the potentials $V$, $V_i$, ${\hat W}_{ij}$, $\cdots$ and their
gradients are computed at the point 1, using the regularization of
Poisson-type integrals defined by formulas like (\ref{6a}). All the
derivatives appearing inside the non-linear sources of the potentials
are considered as distributional and evaluated following the
prescription (\ref{5}). We carefully take into account the fact that
the distributional derivative does not obey the Leibniz rule (it does
satisfy it only in an ``integrated'' sense, thanks to the rule of
integration by parts). An important feature of the equations at the
3PN order is the occurence of some logarithms. From (\ref{6b}) we know
that they are necessarily of the type $\ln(r_{12}/r'_1)$ and
$\ln(r_{12}/s_2)$ in the equations of motion of body 1; interestingly,
the $\ln(r_{12}/s_2)$ appears only in a quartic-interaction term
proportional to $G^4m_1m_2^3$. Thus, at this stage, the 3PN equations
of 1 depend on the constants $\ln r'_1$ and $\ln s_2$ (and {\it idem}
for the equations of 2). Under the form we obtain them, the equations
do not yet admit a conserved energy (of course we are speaking only
about the conservative part of the acceleration, which excludes the
radiation-reaction potential at 2.5PN order). However, we find that a
conserved energy exists if and only if the logarithmic ratios
$\ln(r'_2/s_2)$ and $\ln(r'_1/s_1)$ are adjusted in such a way that

\begin{equation}\label{10}
\ln \left({r'_2\over s_2}\right)=\case{159}{308}
+\lambda~\!{m_1+m_2\over m_2}\quad\mbox{and $~1\leftrightarrow 2$}\;,
\end{equation}
where $\lambda$ is a single numerical constant. If (and only if) the
condition (\ref{10}) is realized, the equations admit an energy and,
in fact, a Lagrangian formulation; in this case, they depend on some
arbitrary constant $\lambda$. The dependence upon the masses in
(\ref{10}) is {\it a priori} allowed. Therefore, the formalism
introduces at this point an undetermined constant $\lambda$. [Using
(\ref{10}), the equations of motion depend also on the constants $\ln
r'_1$ and $\ln r'_2$, but it can be checked that the latter dependence
is pure gauge.]
 
Finally, having in view the application to inspiralling compact
binaries, we present the 3PN relative acceleration and center-of-mass
energy in the case of circular orbits. The acceleration reads as
 
\begin{equation}\label{11}
{d {\bf v}_{12}\over dt} = - \omega^2 {\bf y}_{12} + {1\over c^5} {\bf
F}_{\rm reac} + O\left({1\over c^7}\right)\;,
\end{equation}
where ${\bf y}_{12}={\bf y}_1-{\bf y}_2$ is the relative separation in
harmonic coordinates, ${\bf v}_{12}=d {\bf y}_{12}/dt$ the relative
velocity, and ${\bf F}_{\rm reac}=-\case{32}{5}\case{G^3 m^3
\nu}{r_{12}^4} {\bf v}_{12}$ the standard radiation-reaction force at
the 2.5PN order. The mass parameters are $m=m_1+m_2$, $\mu=\case{m_1
m_2}{m}$ and $\nu=\case{\mu}{m}$. The content of the 3PN
approximation in (\ref{11}) lies in the relation between the
orbital frequency $\omega$ and the coordinate distance $r_{12}=|{\bf
y}_{12}|$. With $\gamma=\case{G m}{r_{12} c^2}$ denoting a small
post-Newtonian parameter, we get

\begin{eqnarray}\label{12}
 \omega^2 = {G m\over r_{12}^3}\biggl\{ 1&+&\left(-3+\nu\right) \gamma
 + \left(6+\case{41}{4}\nu +\nu^2\right) \gamma^2 \\\nonumber
 &+&\left(-10+\left[-\case{67759}{840}+\case{41}{64}\pi^2
+22\ln\left(\case{r_{12}}{r'_0}\right)+\case{44}{3}\lambda
 \right]\nu +\case{19}{2}\nu^2+\nu^3\right) \gamma^3 +O(\gamma^4)
\biggr\}\;.
\end{eqnarray}
The logarithm at 3PN depends on a constant $r'_0$ defined as the
``logarithmic'' barycenter of the two constants $r'_1$ and $r'_2$,
namely $\ln r'_0={m_1\over m}\ln r'_1+{m_2\over m}\ln r'_2$. The
constant $r'_0$ can be eliminated by a change of coordinates. The
center-of-mass energy $E$ of the particles, such that
$\case{dE}{dt}=0$ as a consequence of the conservative equations of
motion (neglecting ${\bf F}_{\rm reac}$), is obtained as
 
\begin{eqnarray}\label{13}
 E = -\case{1}{2}\mu c^2 \gamma\biggl\{1
 &+&\left(-\case{7}{4}+\case{1}{4}\nu\right) \gamma +
 \left(-\case{7}{8}+\case{49}{8}\nu +\case{1}{8}\nu^2\right) \gamma^2
 \\\nonumber
 &+&\left(-\case{235}{64}+\left[\case{106301}{6720}-\case{123}{64}\pi^2
+\case{22}{3}\ln\left(\case{r_{12}}{r'_0}\right)-\case{22}{3}\lambda\right]\nu
 +\case{27}{32}\nu^2+\case{5}{64}\nu^3\right) \gamma^3 +O(\gamma^4)
\biggr\}\;.
\end{eqnarray}
At last, by substituting the expression of $\gamma$ in terms of the
orbital frequency $\omega$ following from the inverse of (\ref{12}),
we find that the 3PN energy in invariant form is given by

\begin{eqnarray}\label{15}
 E = -\case{1}{2}\mu c^2 x
 \biggl\{1&+&\left(-\case{3}{4}-\case{1}{12}\nu\right) x +
 \left(-\case{27}{8}+\case{19}{8}\nu -\case{1}{24}\nu^2\right) x^2
 \\\nonumber
 &+&\left(-\case{675}{64}+\left[\case{209323}{4032}-\case{205}{96}\pi^2
-\case{110}{9}\lambda\right]\nu
 -\case{155}{96}\nu^2-\case{35}{5184}\nu^3\right) x^3 +O(x^4)
\biggr\}\;,
\end{eqnarray}
with $x=(\case{G m \omega}{c^3})^{2/3}$.  In the form (\ref{15}) the
logarithm $\ln (r_{12}/r'_0)$ cancels out. We can compare directly
this result with the one obtained by Jaranowski and Sch\"afer
\cite{JaraS98} (see the equation (5.13) in \cite{DJS1}). We find that
there is perfect agreement provided that $\omega_{\rm
static}=-\case{11}{3}\lambda-\case{1987}{840}$ and $\omega_{\rm
kinetic}=\case{41}{24}$, where $\omega_{\rm static}$ and $\omega_{\rm
kinetic}$ are the two ``ambiguous'' parameters found by Jaranowski and
Sch\"afer. Thus, our undetermined constant $\lambda$ defined by
(\ref{10}) is related to the ambiguous parameter $\omega_{\rm
static}$, while the other ambiguity $\omega_{\rm kinetic}$ takes a
unique value \cite{DJS2}. Let us stress that in the present formalism
we do not meet any ambiguity in the sense of Jaranowski and
Sch\"afer. Rather, the formalism is well-defined thanks in particular
to the rules we employ for handling the pseudo-functions associated
with the Hadamard regularization \cite{BFreg}. All the integrals
encountered in the problem have been given a precise mathematical
sense, and are computed by means of a uniquely defined
prescription. Yet, the appearance of the undetermined constant
$\lambda$ suggests that the present formalism might be physically
incomplete, at least in this present stage of development. Notice that
the constant $\lambda$ enters only the term proportional to
$G^4m_1^2m_2^2(m_1+m_2)$ in the 3PN energy, and that for general
orbits, the energy contains also 164 other terms which are all
uniquely determined. The details of these calculations will be
published elsewhere.

%\acknowledgments

\end{document}